# Statistical equivalence of reduced gravity and enhanced friction in granular packings

Haiyang Lu,[1] Zhikun Zeng,[1] Houfei Yuan,[2] Chengjie Xia,[3] Hanyu Li,[4] Chijin Zhou,[4] Zihang Xu,[4] and Yujie Wang[1,4,5,*]

[1]*School of Physics and Astronomy, Shanghai Jiao Tong University, Shanghai 200240, China*
[2]*Department of Physics, The Hong Kong University of Science and Technology, Hong Kong SAR, China*
[3]*Shanghai Key Laboratory of Magnetic Resonance, School of Physics and Electronic Science, East China Normal University, Shanghai, 200241, China*
[4]*Department of Physics, College of Mathematics and Physics, Chengdu University of Technology, Chengdu 610059, China*
[5]*State Key Laboratory of Geohazard Prevention and Geoenvironment Protection, Chengdu University of Technology, Chengdu 610059, China*

Using X-ray tomography, we compare granular packings prepared under buoyancy-reduced effective gravity with normal gravity packings of particles with systematically varied friction. We show that reducing gravity lowers the random loose packing limit in a manner analogous to increasing friction. Granular packings under reduced gravity and with enhanced friction exhibit identical volume distributions, compactivity, and entropy, indicating that both routes sample statistically equivalent Edwards volume ensembles of mechanically stable states. This equivalence originates from a common relaxation of the mechanical stability constraint: under both conditions, fewer particles are required to participate in the underlying load-bearing bridge structures, leading to a lower contact-number requirement and a higher density of mechanically stable states. Nevertheless, reduced gravity retains a distinct contact-scale signature through more isotropic contact orientations. These findings identify gravity as a physical control governing the statistical accessibility of mechanically stable states within the Edwards framework and provide a unified statistical description of granular packings formed through different physical routes.

Granular materials are athermal systems in which thermal fluctuations are negligible, so their static and slowly driven states are governed by mechanical stability rather than conventional thermal equilibrium [1,2,3]. Nevertheless, the reproducibility of their macroscopic properties under repeated preparation suggests that a statistical description remains possible [4]. This observation motivated Edwards and collaborators to propose an effective statistical-mechanics framework for jammed granular matter, in which volume plays a role analogous to energy and the compactivity $\chi$ serves as a temperature-like variable [3,5]. Within this framework, the volume ensemble is not defined by geometry alone, but is restricted to configurations that satisfy mechanical stability [6]. Friction provides a direct experimental realization of this idea. By introducing tangential contact forces and torque-balance constraints, friction modifies the minimum contact number required for a mechanically stable packing and thereby changes the density of accessible microstates in the Edwards volume ensemble [7,8].

Gravity plays an equally important role in granular materials, but its influence on dense packings remains subtle. Understanding this role is important for terrestrial soils, planetary regolith, and granular processing under reduced-gravity conditions [9-12]. A long line of reduced-gravity experiments using drop towers [13], parabolic flights [14], space-based platforms [15], and buoyancy-controlled sedimentation [16-18] has revealed pronounced changes in granular behavior, including shear strength [19-21], jamming, and glass-like slowing. Among these findings, one of the most striking is that reducing the effective gravity allows mechanically stable packings to exist at lower packing fractions. These observations have generally been interpreted phenomenologically in terms of changes in loading [22,14] or settling dynamics [17]. However, the finding that slow shear and sedimentation in a liquid produce the same limiting packing fraction argues against a purely kinetic explanation based on sedimentation rate or interstitial-fluid viscosity [16]. Instead, it suggests that reduced gravity modifies the conditions for mechanical stability. Although gravity-induced pressure has been proposed as a relevant state variable in a thermodynamic description of the reduced packing fraction [18], whether gravity changes the density of mechanically stable states counted by the Edwards ensemble has never been tested microscopically. This raises a fundamental question: does gravity merely alter how granular packings are prepared and stabilized, or does it reshape the density of accessible mechanically stable configurations?

To address this question, we use X-ray tomography to systematically investigate granular packings under reduced effective gravity and with varying interparticle friction. We show that reducing gravity lowers the random loose-packing (RLP) limit in a manner analogous to increasing friction. Granular packings under reduced gravity and with enhanced friction exhibit equivalent Edwards volume statistics, indicating that both routes sample statistically equivalent ensembles of mechanically stable states. We further show that this equivalence originates from a common reduction in the contact-number constraint, as fewer particles are required to participate in load-bearing bridge structures to maintain mechanical stability. Nevertheless, reduced gravity retains a distinct contact-scale signature by producing more isotropic contact orientations. Our results reveal that gravity enters the Edwards framework not as a passive external field, but as a physical control parameter that regulates the stability constraint and thereby determines the density of accessible mechanically stable states.

We fabricated four bidisperse particle systems, denoted 3DP, BUMP1, BUMP2, and BUMP3, from the same plastic material (VisiJet M2R-WT) using a commercial 3D printer (ProJet MJP 2500 Plus; 0.032 mm resolution). Each system consisted of spheres with diameters $d$ = 3 and 3.6 mm mixed at an approximately equal volume ratio. The 3DP particles had smooth surfaces, whereas BUMP1, BUMP2, and BUMP3 particles were uniformly covered with 500, 300, and 100 hemispherical surface bumps, respectively. Reducing the number of bumps increased the size of the individual asperities and thus the effective surface roughness. Repose-angle measurements showed that the effective friction increased monotonically from 3DP to BUMP3 (see Supplemental Material [23] for details).

Reduced effective gravity was achieved by immersing the particles in an aqueous potassium iodide (KI) solution, which also provided sufficient contrast for X-ray imaging [24]. The effective gravitational acceleration is given by $g_{eff} = \frac{\rho_p - \rho_l}{\rho_p} g = 0.14g$, where $\rho_p = 1.171$ g/cm$^3$ and $\rho_l = 1$ g/cm$^3$ are the particle and liquid densities, respectively. All reduced-gravity results presented below were obtained at $g_{eff} = 0.14g$.

Figure 1(a) shows the experimental setup. Experiments were performed in a cylindrical container (diameter $D = 7$ cm) containing approximately 7500 particles. To suppress crystallization, the inner wall was decorated with randomly distributed ABS hemispheres of diameter 6 mm. To prepare mechanically stable packings under reduced gravity, particles immersed in the KI solution underwent repeated flow-pulse cycles driven by a peristaltic pump (INNOFLUID V6-12L, 0.01 mL resolution). During each cycle, the upward flow fluidized the granular bed for $\tau_F = 1$ s; after the flow ceased, the particles settled under reduced effective gravity for $\tau_W = 5$ s into a mechanically stable packing. Depending on the flow rate $F$ = 30-117 mL/s, 200-5000 pulse cycles were required to reach steady state.

Normal-gravity reference packings were prepared in the same cylindrical container by tapping with a mechanical shaker (DFT9100; 1000 N excitation force). Each tap consisted of a 1/30 s pulse followed by a 1.5 s settling interval. Steady packings were obtained after 10-6000 taps, depending on the tap intensity $\Gamma = 4g$-$12g$.

After reaching steady state, the packings were imaged by desktop X-ray CT. Following the image-processing procedure established in our previous work [25], particle centroids were determined with an error below $3\times10^{-3}d$. The global packing fraction was calculated using $\phi = \sum v_p / \sum v_{\text{voro}}$, where $v_p$ and $v_{\text{voro}}$ are the particle volume and its associated Voronoi cell volume. In the following, only particles that are at least $3d$ away from the boundary of the container are analyzed.

As shown in Fig. 1(b), under normal gravity, the steady-state packing fraction $\phi$ decreases systematically with increasing tapping intensity ($\Gamma / g$) and approaches a lower limiting value for each particle type. We identify this limiting value as the experimentally accessible random loose-packing (RLP) fraction $\phi_{\mathrm{RLP}}$. The measured values are $\phi_{\mathrm{RLP}}$ = 0.604, 0.584, 0.568, and 0.555 for 3DP, BUMP1, BUMP2, and BUMP3, respectively. Increasing friction therefore enables the system to remain mechanically stable at progressively lower packing fractions.

As shown in Fig. 1(c), under the reduced effective gravity of $0.14g$, the steady packing fraction likewise decreases with increasing flow intensity $F$ and approaches the RLP limit. Relative to the normal-gravity results, reduced gravity shifts the RLP of every particle system toward lower packing fractions. In particular, the RLP of 3DP under reduced gravity nearly coincides with that of BUMP1 under normal gravity, while the RLP of BUMP2 under reduced gravity closely matches that of BUMP3 under normal gravity. These constitute two RLP-matched pairs, with $\phi_{\mathrm{RLP}}$ = 0.588 (pair 1) and 0.555 (pair 2). These RLP-matched pairs demonstrate that reduced gravity and increased friction produce nearly identical packing states at the macroscopic level: both enable the granular system to access mechanically stable packings at lower RLP densities.

To determine whether the RLP-matched pairs are also statistically equivalent within the Edwards volume ensemble, we first compare their local-volume distributions $P(V)$ at the corresponding RLP states [see the inset of Fig. 2(a)]. The distributions are nearly identical, and the same local-volume statistics persist for all systems at the same $\phi$, regardless of particle friction or effective gravity. Accordingly, all systems collapse onto a single relationship between the volume variance $\sigma_V^2 = \mathrm{var}(V)$ and $\phi$, as shown in Fig. 2(a). Considering the direct relationship between energy fluctuations and heat capacity in traditional thermodynamic systems [2,26], $\chi$ can be obtained by the integration of volume fluctuations: $\frac{1}{\chi(\phi)}-\frac{1}{\chi^r}=\int_{\phi^r}^{\phi}\frac{d\varphi}{\varphi^2\sigma_V^2}$, where $\phi^r$ is the packing fraction of the reference RLP state with an infinite $\chi^r$. Therefore, the common $\sigma_V^2(\phi)$ relationship naturally leads to the same equation of state $\chi(\phi)$ for each RLP-matched pair, as shown in Fig. 2(b), implying equivalent statistical states.

This common Edwards description is further supported by the observation of the same zeroth-law-like behavior under both normal and reduced gravity. In equilibrium thermodynamics, the zeroth law requires systems coupled to the same thermal bath to reach the same temperature. Analogously, packings composed of different particle types reach the same compactivity $\chi$ when driven at the same intensity, either by tapping [inset of Fig. 2(c)] or by flow-pulse [Fig. 2(c)]. These observations support describing the tapped and flow-pulse packings within a common Edwards volume ensemble rather than treating the reduced-gravity states as purely protocol-dependent dynamical outcomes [27,28]. Following previous studies [7,8], we then calculate the Edwards entropy $S(\phi)-S_{RCP}$ for all systems, taking $S_{RCP}$ as the reference entropy at the random close packing [Fig. 2(d)]. At fixed gravity, increasing friction shifts the entropy curve upward, whereas reducing gravity produces a similar shift for a given particle type. Most importantly, the $S(\phi)$ curves of the two RLP-matched pairs overlap within experimental resolution. Because the Edwards entropy measures the logarithm of the number of mechanically stable configurations, this overlap indicates that each matched pair shares the same density of states.

The remaining differences in the density of states among systems with varying friction coefficients or gravity levels originate from their distinct mechanical-stability requirements built into the Edwards ensemble [6,8,29]. Because the contact number provides a direct measure of the mechanical constraints required for stability, we first examine the average contact number $Z$ in Fig. 3(a). At the same $\phi$, both reduced gravity and increased friction allow mechanically stable packings to exist at lower $Z$, indicating that fewer mechanical constraints are required and, consequently, that a larger number of microscopic configurations become accessible. The decrease in $Z$ is expected to originate from changes in the underlying load-bearing structure. Here, we focus on bridges, cooperative structures in which neighboring particles mutually stabilize one another against an external load such as gravity [30]. Such cooperative structures provide a mode of mechanical stabilization under gravity and are associated with a higher average contact number. Whereas an earlier study reported that bridge structures were largely insensitive to gravity [31], our recent high-resolution X-ray tomography experiments revealed a subtle gravity-induced up-down asymmetry in load-bearing contacts, enabling reliable identification of bridge structures [32]. We therefore examine whether changes in bridge organization account for the reduction in $Z$ under reduced gravity and increased friction.

Following previous studies, particles in mechanically stable packings are classified into singleton particles (SPs), bridge particles (BPs), and unstable particles (UPs) using the lowest-center-of-mass (LCOM) criterion [33,34]. SPs are individually supported by a valid three-particle base, BPs mutually stabilize one another through cooperative supporting structures, and UPs lack a valid supporting base. Figure 3(b) shows the mean contact number $\langle Z_\alpha \rangle$, with $\alpha \in [\mathrm{SP,BP,UP}]$, of the three stability classes in the BUMP2 system as a function of $\phi$. Both $\langle Z_{\mathrm{SP}} \rangle$ and $\langle Z_{\mathrm{BP}} \rangle$ are substantially larger than $\langle Z_{\mathrm{UP}} \rangle$, demonstrating that bridge particles intrinsically require a higher contact number. Notably, the class-resolved contact numbers $\langle Z_\alpha \rangle$ remain nearly unchanged under reduced gravity [Fig. 3(b)], and the inset of Fig. 3(b) shows the same behavior for increasing friction. Thus, neither reduced gravity nor increased friction significantly alters the local connectivity within each stability class.

The principal change instead occurs in the relative populations of the three stability classes. As shown in Fig. 3(c), reducing the effective gravity leaves the singleton fraction ( $f_{\mathrm{SP}}$ ) nearly unchanged, but significantly decreases the bridge-particle fraction ( $f_{\mathrm{BP}}$ ) while increasing the unstable-particle fraction ( $f_{\mathrm{UP}}$ ). A similar redistribution is observed when friction is increased [inset of Fig. 3(c)]. To determine whether these population shifts account for the reduction in the overall contact number, we estimate the overall contact number, $Z_{test} = \sum_{\alpha} \langle Z_{\alpha} \rangle^{1g} \times f_{\alpha}^{0.14g}$ , by fixing the class-resolved contact numbers $\langle Z_{\alpha} \rangle$ at their 1*g* values while replacing only the class fractions $f_{\alpha}$ by those measured at 0.14*g*. As shown by the solid line in Fig. 3(a), the close agreement between $Z_{test}$ and the measured *Z* at 0.14*g* for all frictional systems demonstrates that the reduction in *Z* is primarily caused by the redistribution among the three stability classes. Under normal gravity or lower friction, a larger fraction of particles must participate in bridge structures to satisfy mechanical stability. Because bridge particles are associated with a higher average contact number, their larger population under normal gravity or lower friction naturally gives rise to a higher overall contact number *Z* at the same packing fraction $\phi$ .

To further understand how the contact-number constraint governs the Edwards density of states, we examine the relationship between the contact number *Z* and the Edwards entropy *S*. As shown in Fig. 3(d), all datasets collapse onto a single linear *S*(*Z*) relation irrespective of gravity level or friction coefficient. This collapse demonstrates that the Edwards entropy (or equivalently, the density of mechanically stable states) is governed primarily by the contact-number constraint, consistent with the mean-field prediction proposed by Makse et al. [35,36]. Physically, increasing the contact number imposes additional mechanical constraints, thereby reducing the configurational degrees of freedom available to mechanically stable packings and consequently the Edwards density of states. Our measurements therefore provide direct experimental evidence that gravity and friction regulate the Edwards density of states through the common microscopic mechanism of the contact-number constraint.

The results above establish that reduced gravity and enhanced friction become statistically equivalent within the Edwards volume ensemble at the particle scale. This naturally raises the question of whether the two systems also have the same directional organization of the contact network. Previous studies have shown that different preparation protocols can generate distinct contact anisotropies even when the resulting packings exhibit nearly identical volume statistics [8,37-39]. We therefore compare the contact orientation statistics of the two RLP-matched pairs. As shown in Fig. 4(a), we measure the probability distribution PDF( $\theta$ ) of the polar angle between each contact vector and the vertical direction for the two RLP-matched pairs, with pair 1 at $\phi$ = 0.59 and pair 2 at $\phi$ = 0.57. The normal-gravity packings exhibit a stronger angular modulation, whereas the reduced-gravity packings display more uniform distributions, indicating a more isotropic distribution of contact orientations. This difference is quantified by the *z*-component of the contact fabric tensor $|A_{zz}|$ (see Supplemental Material [23] for details). As shown in the insets of Fig. 4(a), both RLP-matched pairs exhibit smaller $|A_{zz}|$ under reduced gravity, confirming that reduced gravity weakens the directional bias of contact orientations. Thus, despite the Edwards states equivalence, gravity still retains a distinct contact-scale mechanical signature through anisotropy in contact orientations.

To understand why the Edwards volume statistics remain identical despite the different contact anisotropies, we examine the geometry of the Voronoi tessellation, which directly determines the Edwards volume ensemble. We first calculate the scalar Voronoi-cell anisotropy $\langle \beta_{lcoal} \rangle$ from the rank-2 Minkowski tensor $\mathbf{W}_0^{0,2}$ (see Supplemental Material [23] for details). As shown in Fig. 4(b), $\langle \beta_{lcoal} \rangle$ remains nearly identical for all systems, indicating that the particle-scale geometry is essentially unchanged. We then analyze the orientation of the Voronoi cells through the distribution PDF( $\varphi$ ), where $\varphi$ is the angle between the principal axis of each Voronoi cell and the vertical direction. As shown in Fig.4(c), the reduced gravity packings exhibit a more uniform orientation distribution, corresponding to a smaller normalized z-component of the tensor $|\langle W_{zz} \rangle|$ [inset of Fig. 4(c)]. These results indicate that gravity primarily affects the preferred orientation of the Voronoi cells rather than their intrinsic geometry.

These observations reveal a clear separation between the responses of contact-scale and particle-scale structures to gravity. This explains why reduced gravity and enhanced friction become statistically equivalent within the Edwards framework despite retaining distinct contact-scale mechanical signatures. These results demonstrate that this equivalence is governed primarily by the average mechanical constraint encoded by the contact number $Z$, rather than by the detailed anisotropic organization of the contact orientations.

Taken together, our X-ray tomography measurements reveal that reduced gravity and enhanced friction represent two distinct physical routes to statistically equivalent ensembles of mechanically stable states within the Edwards framework. This equivalence arises from a common relaxation of the mechanical-stability constraint: under both conditions, fewer particles are required to participate in bridge structures, reducing the contact-number requirement and thereby increasing the density of mechanically stable states. Reduced gravity nevertheless retains a distinct signature at the contact scale through more isotropic contact orientations, showing that this equivalence is confined to the particle-scale Edwards description and does not extend to the finer contact-scale organization.

From this perspective, gravity should be viewed not merely as an external loading condition that governs the local force balance and dynamical evolution of a granular assembly, but as one of the physical controls that determine how mechanically stable states are statistically sampled. Consequently, the influence of gravity is naturally incorporated into the Edwards framework through its effect on the mechanical stability constraint. Our findings therefore suggest that the Edwards framework provides a unified statistical description of granular packings formed through different physical routes, offering a physical basis for extending this picture to broader preparation conditions. Such a perspective may prove valuable for understanding buoyancy-controlled sediments, seabed stability, planetary regolith, and granular materials under reduced or microgravity environments.

The work has been supported by the National Natural Science Foundation of China (No. 12534008, 12274292) and the Space Application System of China Manned Space Program (No. KJZ-YY-NLT0504). Z.Z. acknowledges support from the National Natural Science Foundation of China (No. 123B2060).

Corresponding author

*yujiewang@sjtu.edu.cn

## Reference

[1] H. M. Jaeger, S. R. Nagel, and R. P. Behringer, Granular solids, liquids, and gases, Rev. Mod. Phys. **68**, 1259 (1996).
[2] D. P. Bi, S. Henkes, K. E. Daniels, and B. Chakraborty, The Statistical Physics of Athermal Materials, Annu. Rev. Condens. Matter Phys. **6**, 63 (2015).
[3] S. F. Edwards and R. B. S. Oakeshott, Theory of Powders, Physica A **157**, 1080 (1989).
[4] E. R. Nowak, J. B. Knight, E. Ben-Naim, H. M. Jaeger, and S. R. Nagel, Density fluctuations in vibrated granular materials, Phys. Rev. E **57**, 1971 (1998).
[5] A. Mehta and S. F. Edwards, Statistical-Mechanics of Powder Mixtures, Physica A **157**, 1091 (1989).
[6] C. Song, P. Wang, and H. A. Makse, A phase diagram for jammed matter, Nature **453**, 629 (2008).
[7] Y. Yuan *et al.*, Experimental Test of the Edwards Volume Ensemble for Tapped Granular Packings, Phys. Rev. Lett. **127**, 018002 (2021).
[8] Y. Xing, Y. Yuan, H. F. Yuan, S. Y. Zhang, Z. K. Zeng, X. Zheng, C. J. Xia, and Y. J. Wang, Origin of the critical state in sheared granular materials, Nat. Phys. **20**, 646 (2024).
[9] N. Murdoch, B. Rozitis, S. F. Green, T. L. de Lophem, P. Michel, and W. Losert, Granular shear flow in varying gravitational environments, Granul. Matter **15**, 129 (2013).
[10] N. Murdoch, B. Rozitis, S. F. Green, P. Michel, T. L. de Lophem, and W. Losert, Simulating regoliths in microgravity, Mon. Not. R. Astron. Soc. **433**, 506 (2013).
[11] J. Brisset, J. Colwell, A. Dove, S. Abukhalil, C. Cox, and N. Mohammed, Regolith behavior under asteroid-level gravity conditions: low-velocity impact experiments, Prog. Earth Planet. Sci. **5**, 73 (2018).
[12] D. J. Scheeres, C. M. Hartzell, P. Sánchez, and M. Swift, Scaling forces to asteroid surfaces: The role of cohesion, Icarus **210**, 968 (2010).
[13] O. D'Angelo, Q. Yu, and T. Pöschel, Granular jamming and rheology in microgravity, J. Rheol. **70**, 501 (2026).
[14] O. D'Angelo, A. Horb, A. Cowley, M. Sperl, and W. T. Kranz, Granular piston-probing in microgravity: powder compression, from densification to jamming, npj Microgravity **8**, 48 (2022).
[15] C. Mayo, M. Kunzner, M. Sperl, and J. P. Gabriel, Observing the Glass and Jamming Transitions of

Dense Granular Material in Microgravity, Phys. Rev. Lett. **135,** 178203 (2025).
[16] G. Y. Onoda and E. G. Liniger, Random Loose Packings of Uniform Spheres and the Dilatancy Onset, Phys. Rev. Lett. **64**, 2727 (1990).
[17] G. R. Farrell, K. M. Martini, and N. Menon, Loose packings of frictional spheres, Soft Matter **6**, 2925 (2010).
[18] M. Jerkins, M. Schröter, H. L. Swinney, T. J. Senden, M. Saadatfar, and T. Aste, Onset of mechanical stability in random packings of frictional spheres, Phys. Rev. Lett. **101**, 018301 (2008).
[19] X. H. Cheng, S. Z. Xiao, S. Yang, N. F. Zhao, and A. S. Cao, A pressure-sensitive rheological origin of high friction angles of granular matter observed in NASA-MGM project, Chin. Phys. B **33**, 068301 (2024).
[20] M. G. Kleinhans, H. Markies, S. J. de Vet, A. C. I. Veld, and F. N. Postema, Static and dynamic angles of repose in loose granular materials under reduced gravity, J. Geophys. Res. Planets **116**, E11004 (2011).
[21] J. P. Marshall, R. C. Hurley, D. Arthur, I. Vlahinic, C. Senatore, K. Iagnemma, B. Trease, and J. E. Andrade, Failures in sand in reduced gravity environments, J. Mech. Phys. Solids **113**, 1-12 (2018).
[22] N. Murdoch, B. Rozitis, K. Nordstrom, S. F. Green, P. Michel, T. L. de Lophem, and W. Losert, Granular Convection in Microgravity, Phys. Rev. Lett. **110**, 018307 (2013).
[23] See Supplemental Material for details on repose angle measurement, basic packing characteristics and calculation details.
[24] Y. Xing, Y. P. Qiu, Z. Wang, J. C. Ye, and X. T. Li, X-ray tomography study on the structure of the granular random loose packing, Chin. Phys. B **26**, 084503 (2017).
[25] Z. F. Li, Z. K. Zeng, Y. Xing, J. D. Li, J. Zheng, Q. H. Mao, J. Zhang, M. Y. Hou, and Y. J. Wang, Microscopic structure and dynamics study of granular segregation mechanism by cyclic shear, Science Advances **7**, eabe8737 (2021).
[26] C. J. Xia, J. D. Li, Y. X. Cao, B. Q. Kou, X. H. Xiao, K. Fezzaa, T. Q. Xiao, and Y. J. Wang, The structural origin of the hard-sphere glass transition in granular packing, Nat. Commun. **6**, 8409 (2015).
[27] M. P. Ciamarra, A. Coniglio, and M. Nicodemi, Thermodynamics and statistical mechanics of dense granular media, Phys. Rev. Lett. **97**, 158001 (2006).
[28] M. Schröter, D. I. Goldman, and H. L. Swinney, Stationary state volume fluctuations in a granular medium, Phys. Rev. E **71**, 030301 (2005).
[29] C. Briscoe, C. Song, P. Wang, and H. A. Makse, Entropy of Jammed Matter, Phys. Rev. Lett. **101**, 188001 (2008).
[30] A. Mehta, G. C. Barker, and J. M. Luck, Heterogeneities in granular materials, Phys. Today **62**, 40 (2009).
[31] M. C. Jenkins, M. D. Haw, G. C. Barker, W. C. K. Poon, and S. U. Egelhaaf, Does Gravity Cause Load-Bearing Bridges in Colloidal and Granular Systems?, Phys. Rev. Lett. **107**, 038302 (2011).
[32] C. J. Zhou, S. Y. Zhang, X. L. Dai, Y. X. Cao, Y. Yuan, C. J. Xia, Z. K. Zeng, and Y. J. Wang, Identifying bridges from asymmetric load-bearing structures in tapped granular packings, Commun. Phys. **8**, 312 (2025).
[33] Y. X. Cao, B. Chakrabortty, G. C. Barker, A. Mehta, and Y. J. Wang, Bridges in three-dimensional granular packings: Experiments and simulations, Europhys. Lett. **102**, 24004 (2013).
[34] A. Mehta, G. C. Barker, and J. M. Luck, Cooperativity in sandpiles: statistics of bridge geometries, J. Stat. Mech. Theory Exp. **2004**, P10014 (2004).
[35] A. Baule, F. Morone, H. J. Herrmann, and H. A. Makse, Edwards statistical mechanics for jammed granular matter, Rev. Mod. Phys. **90**, 015006 (2018).
[36] P. Wang, C. M. Song, Y. L. Jin, and H. A. Makse, Jamming II: Edwards' statistical mechanics of random packings of hard spheres, Physica A **390**, 427 (2011).
[37] D. P. Bi, J. Zhang, B. Chakraborty, and R. P. Behringer, Jamming by shear, Nature **480**, 355 (2011).
[38] J. Sun and S. Sundaresan, A constitutive model with microstructure evolution for flow of rate-independent granular materials, J. Fluid Mech. **682**, 590 (2011).
[39] H. Y. Li, H. F. Yuan, Z. K. Zeng, S. Y. Zhang, C. J. Zhou, X. Y. Ai, and Y. J. Wang, Microscopic structural study on the growth history of granular heaps prepared by the raining method, Soft Matter **21**, 7565 (2025).

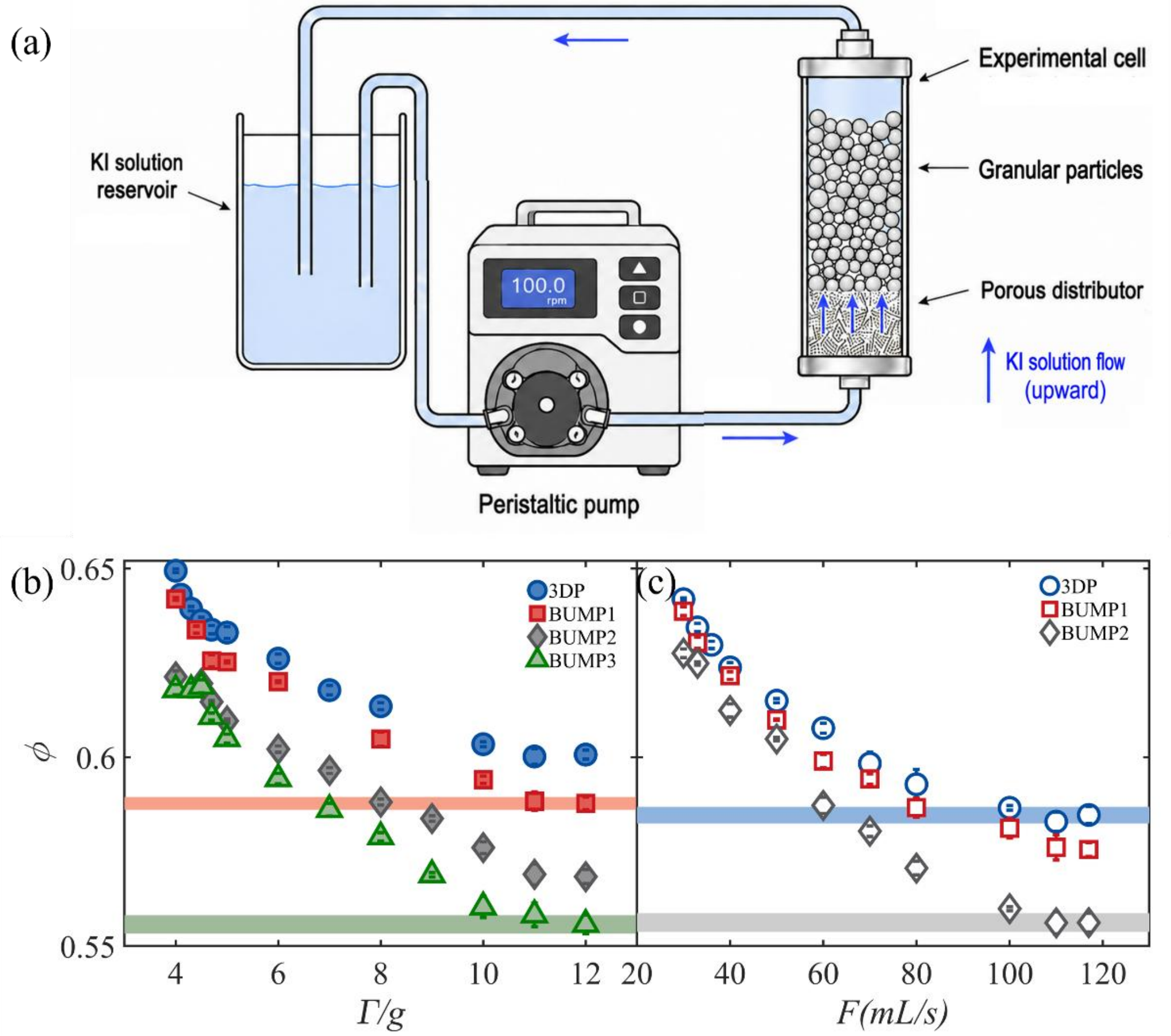

FIG. 1. (a) Schematic of the experimental setup. Packing fraction $\phi$ as a function of (b) tap intensity $\Gamma / g$ for the four particle systems at normal gravity, 1$g$, and (c) flow-pulse intensity $F$ (mL/s) for the three particle systems at reduced gravity, 0.14$g$. The solid horizontal lines in (b) and (c) mark $\phi$ of the two RLP-matched pairs discussed in the text.

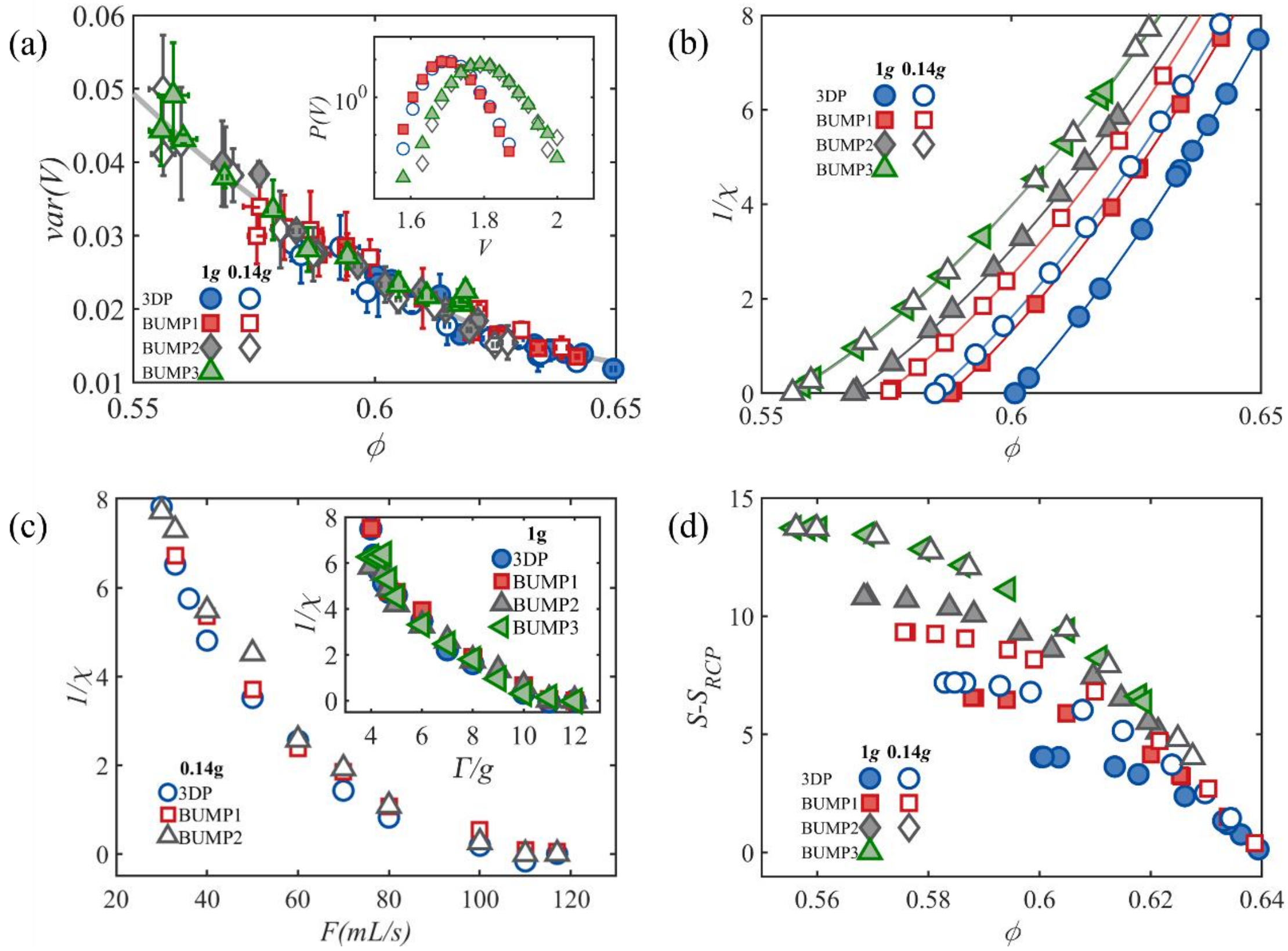

FIG. 2. (a) Volume fluctuation var($V$) as a function of $\phi$ (symbols); the solid curve is a quadratic polynomial fit. Inset: probability distributions $P(V)$ for four RLP states in the two RLP-matched pairs. (b) Inverse compactivity $\chi^{-1}$ as a function of $\phi$ for systems at 0.14$g$ (open symbols) and 1$g$ (solid symbols). (c) Inverse compactivity $\chi^{-1}$ as a function of flow-pulse intensity $F$ (mL/s). Inset: $\chi^{-1}$ as a function of tap intensity $\Gamma / g$. (d) Edwards entropy $S$ as a function of $\phi$.

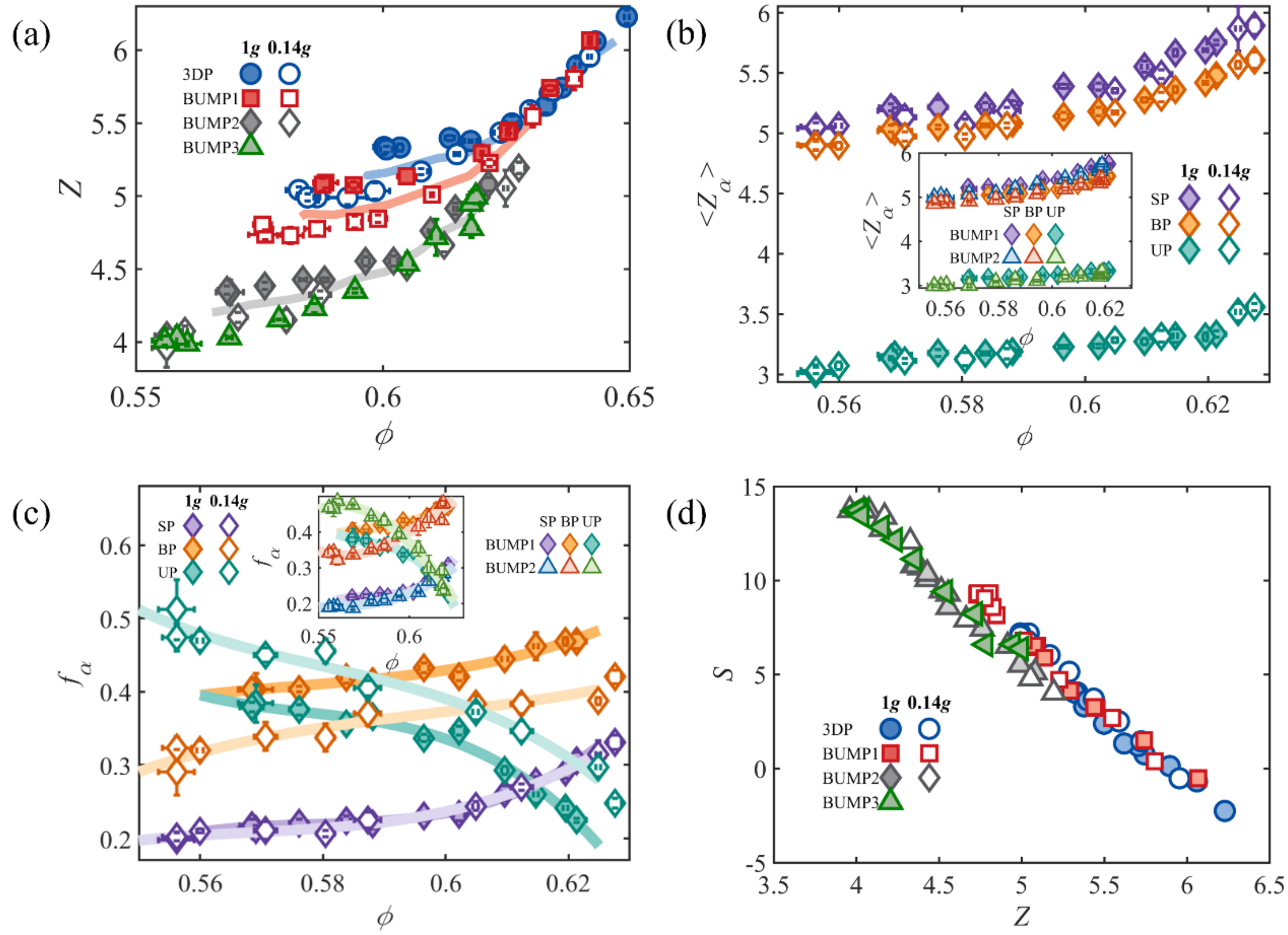

FIG. 3. (a) Average contact number $Z$ as a function of $\phi$ for systems at 0.14$g$ (open symbols) and 1$g$ (solid symbols). The solid curves show $Z_{test}$ as a function of $\phi$ (see the main text). (b) Mean contact number $\langle Z_{\alpha} \rangle$, with $\alpha \in$[SP,BP,UP] of the three stability classes, as a function of $\phi$. (c) Fractions $f_{\alpha}$ as functions of $\phi$. Both panels of (b) and (c): BUMP2 at 1$g$ (filled symbols) and 0.14$g$ (open symbols). Insets of both (b) and (c): BUMP2 (diamonds) and BUMP3 system (triangles) at 1$g$. (d) Edwards entropy $S$ as a function of $Z$ for all systems.

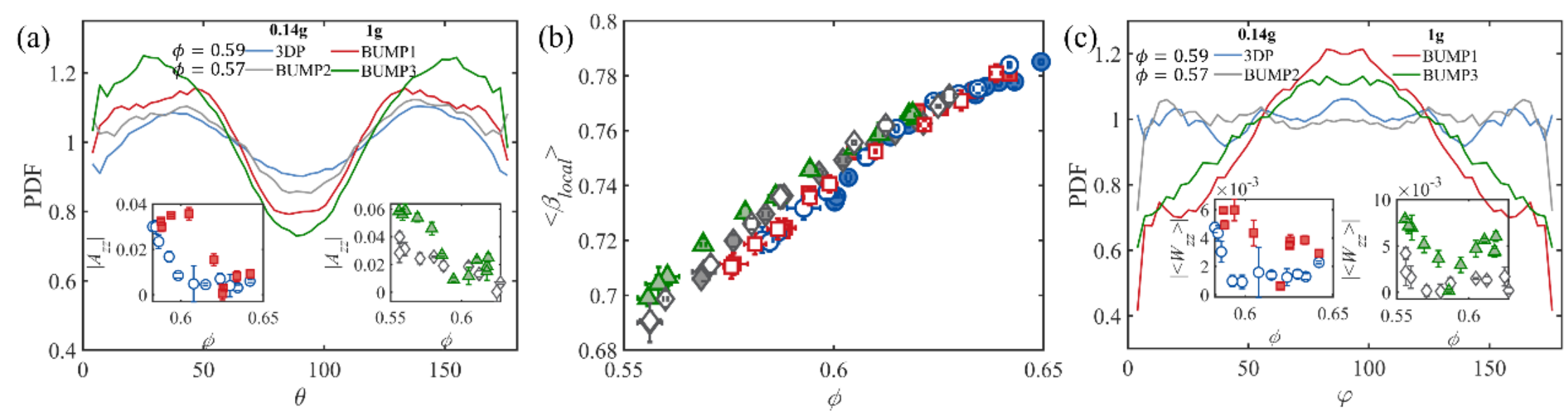

FIG. 4. (a) Contact-orientation distributions PDF( $\theta$ ) for the two RLP-matched pairs, at $\phi$ = 0.59 and $\phi$ = 0.57, respectively. Insets: z-component of fabric tensor $|A_{zz}|$, as a function of $\phi$ for pair 1 (left inset) and pair 2 (right inset). (b) Voronoi-cell anisotropy $\langle \beta_{local} \rangle$ as a function of $\phi$. (c) Voronoi-cell principal-axis orientation distributions PDF( $\varphi$ ) as functions of $\varphi$ for the same systems as in (a). Insets: z-component of the rank-2 Minkowski tensor $|\langle W_{zz} \rangle|$ as a function of $\phi$ for pair 1 (left) and pair 2 (right).

# Supplemental Material for

# Statistical equivalence of reduced gravity and enhanced friction in granular packings

Haiyang Lu,[1] Zhikun Zeng,[1] Houfei Yuan,[2] Chengjie Xia,[3] Hanyu Li,[4] Chijin Zhou,[4] Zihang Xu,[4] and Yujie Wang[1,4,5,*]

*[1]School of Physics and Astronomy, Shanghai Jiao Tong University, Shanghai 200240, China*
*[2]Department of Physics, The Hong Kong University of Science and Technology, Hong Kong SAR, China*
*[3]Shanghai Key Laboratory of Magnetic Resonance, School of Physics and Electronic Science, East China Normal University, Shanghai, 200241, China*
*[4]Department of Physics, College of Mathematics and Physics, Chengdu University of Technology, Chengdu 610059, China*
*[5]State Key Laboratory of Geohazard Prevention and Geoenvironment Protection, Chengdu University of Technology, Chengdu 610059, China*

Corresponding author

*yujiewang@sjtu.edu.cn

## 1. Repose angle measurement

We measure the effective friction coefficients for 3DP, BUMP1, BUMP2 and BUMP3 beads under normal gravity (1*g*) and buoyancy reduced effective gravity (0.14*g*) by the repose angle method. Beads are fed into a rotating drum with inner diameter 11 cm and width 5 cm, occupying nearly half of its inner volume. Under continuous quasistatic rotation, the system displays increasing unsteady slopes accompanied by avalanches of different magnitudes. We record this process by video imaging and measure the repose angle $\theta$ just after each avalanche. Thus, we can obtain $\mu = \tan\theta$ by averaging over 100 independent realizations with $\mu_{\mathrm{3DP}}^{1g} = 0.481$, $\mu_{\mathrm{BUMP1}}^{1g} =$ 0.520, $\mu_{\mathrm{BUMP2}}^{1g} = 0.596$, $\mu_{\mathrm{BUMP3}}^{1g} = 0.669$, $\mu_{\mathrm{3DP}}^{0.14g} = 0.517$, $\mu_{\mathrm{BUMP1}}^{0.14g} = 0.572$ and $\mu_{\mathrm{BUMP2}}^{0.14g} = 0.63$, as shown in Fig. S1. Figure S1 shows that $\mu$ monotonically decreases with increasing packing fraction $\phi$, irrespective of the gravity level.

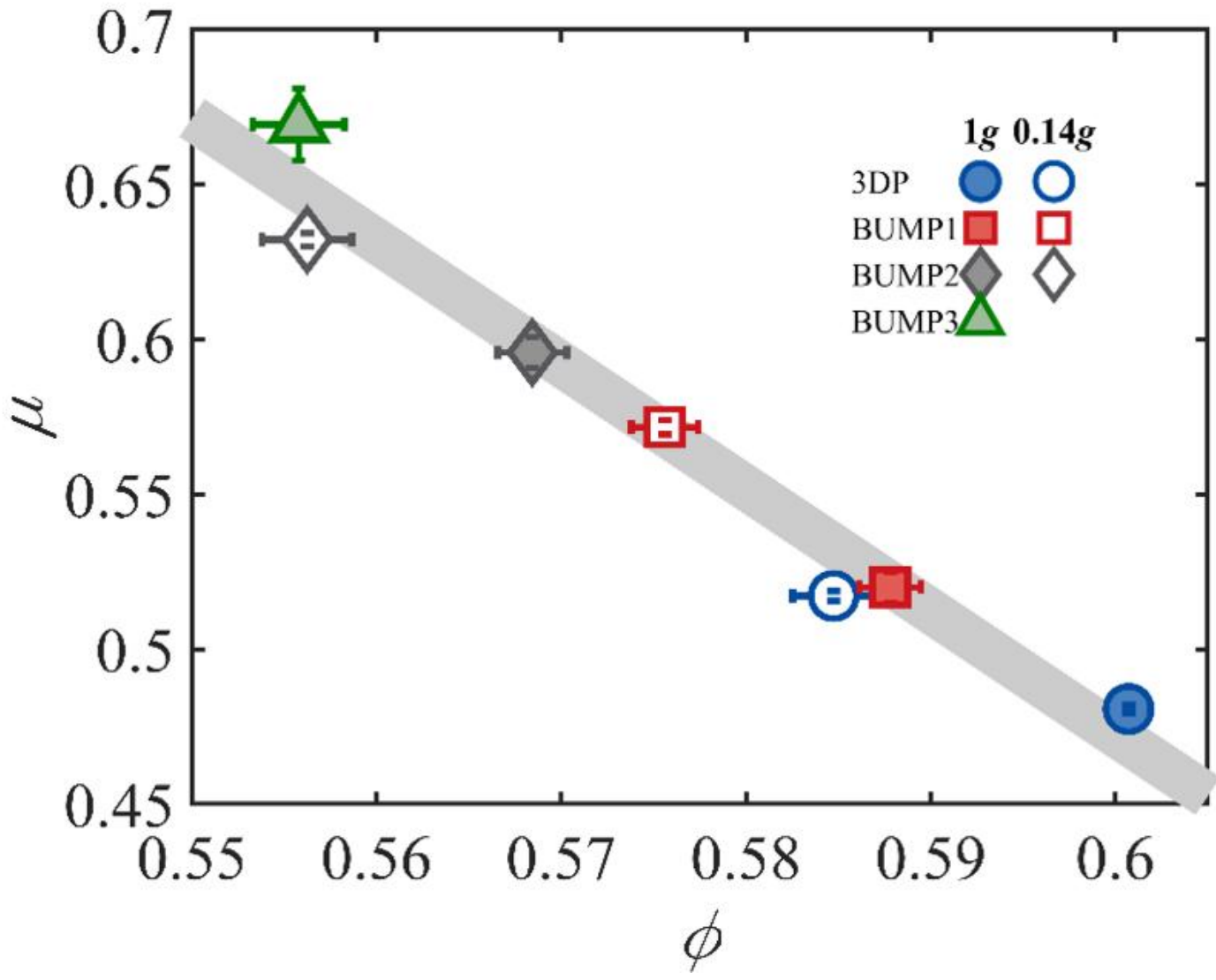

FIG. S1. Effective friction coefficients $\mu$ as a function of packing fraction $\phi$ for both normal gravity 1*g* (filled symbols) and reduced gravity 0.14*g* (open symbols)

**2. Initial-state preparation**

Before each data series, a reproducible initial state was prepared by applying 50 cycles at the strongest driving condition: 50 flow pulses at $F$ = 117 mL/s for the reduced-gravity experiments and 50 taps at $\Gamma$ = 12*g* for the normal-gravity experiments.

**3. Basic packing characteristics**

Although ordering close to the container boundary is avoided by the introduced roughness of the wall (see main text), the Voronoi tessellation still gives relatively large cell volumes for boundary particles (including those close to the free surface). Therefore, we exclude particles within 3d from the boundary for all the following analyses. For the remaining particles, we measure the local packing fraction $\phi_{local}$ binned along the vertical direction for both normal gravity and reduced effective gravity systems, as shown in Figs. S2a and S2b. The variation of $\phi_{local}$ for a certain system is within about 0.01, i.e., no apparent gradient is observed. Thus, local packing structures are nearly homogeneous after excluding the boundary particles.

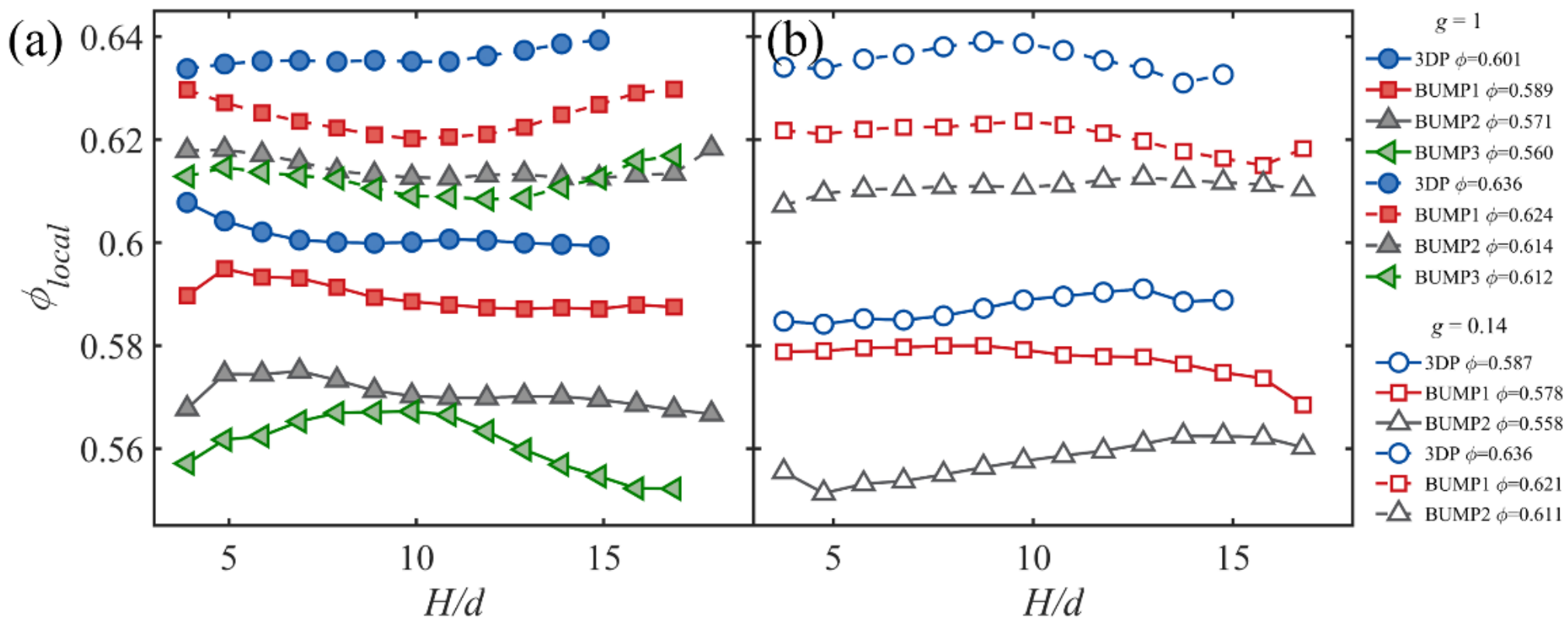

FIG. S2. Local packing fraction $\phi_{local}$ measured in different bins along the vertical direction *H* from the bottom for both normal gravity 1*g* (a) and reduced gravity 0.14*g* (b) systems at different (global) $\phi$.

## 4. Calculation details

The fabric tensor is defined as $\mathbf{A} = \frac{1}{N}\sum_{a=1}^{N_c} \mathbf{n}^a \otimes \mathbf{n}^a - \frac{1}{3}\mathbf{I}$, where $N_c$ is the total number of contacts, $\mathbf{n}^a$ is the unit contact vector between the centers of two contacting particles, and I is the unit tensor. $|A_{zz}|$ is the absolute value of the z-direction component of $\mathbf{A}$.

The rank-2 Minkowski tensor $\mathbf{W}_0^{0,2}$ of each Voronoi cell is defined as $\mathbf{W}_0^{2,0} = \int \mathbf{r} \otimes \mathbf{r} dV$, where r is the point vector connecting the particle center and each volume element. The Voronoi cell anisotropy is characterized by the ratio of the maximum and minimum eigenvalues of $\mathbf{W}_0^{0,2}$, denoted by $\beta_{local}$. $\langle \beta_{lcoal} \rangle$ is an average of $\beta_{local}$ among all constituting particles. The orientation of each Voronoi cell was identified as the eigenvector with the largest eigenvalue of $\mathbf{W}_0^{0,2}$. $|\langle W_{zz} \rangle|$ is the normalized z-direction component of the rank-2 Minkowski tensor and the absolute value is taken after averaging over all analyzed particles.